\documentclass[%
 amsmath,amssymb,
 aps, 
 prl,
 10pt,
superscriptaddress,
]{revtex4-2}

\usepackage{braket}%
\usepackage{graphicx}%
\usepackage{multirow}%
\usepackage{amsmath,amssymb,amsfonts}%
\usepackage{amsthm}%
\usepackage{mathrsfs}%
\usepackage[version=4]{mhchem}
\usepackage[nolist]{acronym}
\usepackage{xr}
\usepackage{siunitx}
\usepackage{pdfpages}
\makeatletter
\AtBeginDocument{\let\LS@rot\@undefined}
\makeatother

\begin{document}

\preprint{APS/123-QED}

\title{\textbf{Multi-fidelity Machine Learning Interatomic Potentials for Charged Point Defects} 
}%

\author{Xinwei Wang}
\email{xinwei.wang20@imperial.ac.uk}
\affiliation{Thomas Young Centre and Department of Materials, Imperial College London, London SW7 2AZ, UK}
\affiliation{Imperial-X, Imperial College London, London W12 0BZ, UK}

\author{Irea Mosquera-Lois}%
\affiliation{Thomas Young Centre and Department of Materials, Imperial College London, London SW7 2AZ, UK}

\author{Aron Walsh}
\affiliation{Thomas Young Centre and Department of Materials, Imperial College London, London SW7 2AZ, UK}

\date{\today}

\begin{acronym}
\acro{MLIPs}{Machine learning interatomic potentials}
\acro{MLIP}{machine learning interatomic potential}
\acro{PESs}{potential energy surfaces}
\acro{PES}{potential energy surface}
\acro{DFT}{density functional theory}
\acro{RMSD}{root mean square deviation}
\acro{PCA}{principal component analysis}
\acro{MF}{multi-fidelity}
\acro{HF}{high-fidelity}
\acro{LF}{low-fidelity}
\acro{CCDs}{configuration coordinate diagrams}
\acro{CCD}{configuration coordinate diagram}
\acro{2D}{two-dimensional}
\end{acronym}

\begin{abstract}
Machine learning interatomic potentials (MLIPs) can now reproduce the energy, forces and stresses of bulk materials with high accuracy compared to first-principles calculations. The description of imperfections, where coordination environments and electron counts deviate from those found in pristine reference structures, remains a challenge. We find that the current generation of foundation MLIPs do not describe the defect physics of the semiconductor \ce{Sb2Se3}. We introduce global defect charge embeddings that distinguish the bonding characteristics of different charge states. We further employ a multi-fidelity approach that combines low-cost (semi-local exchange-correlation functional) reference data with high-quality (non-local hybrid functional) energies and forces that describe well the subtleties of the defect energy landscape. The resulting defect-capable force fields can find stable structural configurations and predict defect thermodynamics in quantitative agreement with direct quantum mechanical calculations, at a fraction of the computational cost.  
\end{abstract}

\maketitle



Defects play a crucial role in determining the properties of functional materials, governing key processes in semiconductors, photovoltaics, solid-state electrolytes, and catalytic systems \cite{freysoldt2014first,walsh2017instilling}. 
Accurate predictions of these properties rely fundamentally on identifying the correct ground-state atomic structure.
Failure to locate the equilibrium configuration impacts subsequent calculations of formation energies, migration barriers, carrier capture rates and degradation reactions \cite{mosquera-lois_search_2021,kavanagh2022impact,wang2023fourelectron,el2004sampling,Squires2024}. Ground and metastable structures can be identified through structure searching methods that explore the energy landscape via symmetry-breaking distortions \cite{mosquera2022shakenbreak}. 
However, comprehensive structure searching remains a major bottleneck \cite{pickard2011ab}, especially for systems with complex defect landscapes where multiple metastable configurations compete within a narrow energy window \cite{wang2024sulfur}.
Accurately describing these landscapes requires a high level of electronic-structure theory, such as hybrid functionals, but their computational cost renders exhaustive structure searching prohibitively expensive.
To reduce the cost of the underlying calculations, current approaches rely on low-precision approximations (e.g. coarse sampling of reciprocal space), which do not ensure identification of the true global minimum.

\ac{MLIPs} offer a promising route to address this bottleneck.
Recent foundation models \cite{batatia2025crosslearning,batatia2025foundation,lysogorskiy2025graph,yang2024mattersim} trained on comprehensive datasets \cite{jain2013commentary,barroso_omat24,schmidt2024improving} have demonstrated generalization across chemical space for pristine systems, with prediction accuracy approaching quantum mechanical calculations \cite{loew2025universal}.
However, their application to defect systems remains an open challenge.
First, \ac{MLIPs}-based defect modeling faces a data scarcity problem more severe than bulk materials.
Defects require training datasets that capture a wide range of local atomic environments, including symmetry-breaking distortions and meta-stable configurations, all calculated at a high level of theory.
Charged defects, which govern carrier recombination and doping behavior in semiconductors \cite{freysoldt2014first}, pose an additional challenge.
Different charge states of the same defect can have distinct ground-state geometries because changes in electronic occupation modify the local bonding.
However, standard \ac{MLIPs} are built on the locality assumption \cite{prodan2005nearsightedness}, using descriptors that capture only the chemical species and local atomic environment within a truncated cutoff \cite{behler2007generalized,batatia2022mace}.
This representation contains no explicit information about the charge state, so configurations with similar local environments produce nearly identical descriptors despite corresponding to distinct charge states, thereby mapping the multiple \ac{PESs} into a single energy landscape.
Consequently, existing studies largely focus on neutral defects \cite{mannodi2022universal, mosquera2024machine,kavanagh2024identifying,zhou2025one,yang2025modeling,turiansky2025machinelearningphononspectra}, although these are not intrinsically simpler for \ac{MLIP} modeling, as they often undergo substantial structural reconstructions driven by excess electrons or holes localized on the defect.
Others train separate models for each charge state \cite{mosquera2025point,mosquera2025dynamic,rahman2025defect}, rather than a single model that distinguishes charge states within a unified architecture.
Despite growing developments in charge-aware ML models, their application to charged defects in periodic semiconductor supercells remains limited.
Methods that explicitly model charge redistribution and long-range electrostatics, such as learned long-range descriptors \cite{cheng2025latent}, charge equilibration schemes \cite{ko2021fourth,vondrak2026integrating} and polarizable electrostatic models \cite{batatia2026mace,baldwin2026design}, have been developed primarily for bulk liquids, molecular crystals and interfaces.
However, for charged defects, these approaches can suffer from ill-defined atomic charge partitioning \cite{wang2014modeling}, unphysical delocalization into the bulk \cite {vondrak2026integrating} or the non-linear scaling of Ewald summation \cite{toukmaji1996ewald}.
Global charge conditioning, which encodes the total system charge through a learnable embedding \cite{wood2025family}, offers a simpler alternative that captures charge information and distinguishes charge states with negligible additional computational cost or complexity.
Across both approaches, a single model capable of simultaneously distinguishing multiple charge states and subtle energy differences between competing defect configurations has not yet been demonstrated \cite{shimizu2022using,kiyohara2025machine,wood2025family}.

In this Letter, we demonstrate that:
(i) foundation machine learning force fields, trained on bulk materials data, do not reliably identify defect ground states and often miss low-energy configurations;
(ii) bespoke models with global defect charge embeddings can successfully identify ground-state configurations to within 0.05 Å, and accurately predict defect formation energies and charge transition levels to within 0.02 eV;
(iii) a \ac{MF} approach that combines a high volume of low-cost reference (PBE) data with selective high-quality (HSE) data results in a robust model that can identify global minima that are missed in low-precision searches.
The resulting defect-capable force fields act as fast surrogate models that can describe dynamic processes that are beyond the limits of traditional defect modeling approaches. The \ac{MF} strategy with charge embeddings also provides a solid foundation towards the development of foundation models for defective crystals. 

\begin{figure*}[ht]
    \centering    {\includegraphics[width=\textwidth]{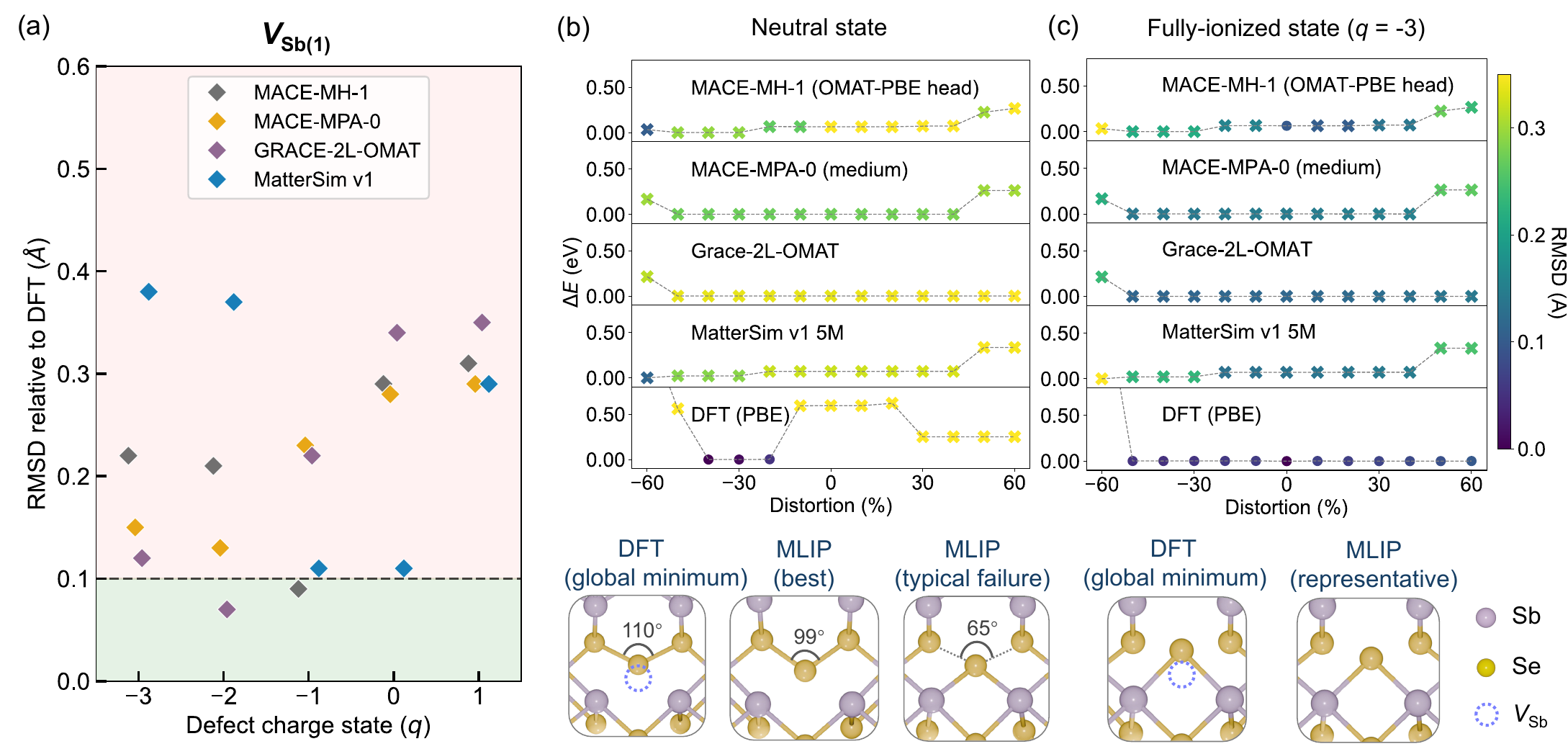}} \\
    \caption{Failure of foundation machine learning interatomic potentials (MLIPs) to capture defect structures.
    (a) Root mean square deviation (RMSD) of MLIP-predicted ground-state structures relative to density functional theory (DFT) calculations using the PBE functional for $V_{\text{Sb(1)}}$ in \ce{Sb2Se3} across five charge states. For each charge state, data points are slightly offset along the horizontal axis for clarity. The green shaded region (RMSD $\leq$ 0.1 Å) indicates successful structural identification.
    (b, c) Comparison of defect potential energy surfaces (PESs) and structural predictions for (b) the neutral ($q=0$) and (c) fully ionized ($q=-3$) charge states of $V_{\text{Sb(1)}}$. The PESs are mapped as a function of the bond distortion percentage, with energies referenced to the minimum of each respective landscape. In these panels, data points for the foundation models and the DFT reference are colored according to the RMSD of the relaxed atomic positions relative to the DFT reference ground-state, with circles indicating RMSD $\leq$ 0.1 Å and crosses denoting RMSD $>0.1$ Å. Bottom insets compare the local atomic environment of the DFT global minimum with representative MLIP-predicted minima.
    }
    \label{fig_1}
\end{figure*}

\textbf{Limitation of foundation models for charged defects} — 
Foundation models trained on massive bulk datasets have enabled rapid property prediction for crystalline materials, yet their applicability to the complex landscape of charged point defects remains unexplored.
To address this gap, we first benchmarked four state-of-the-art equivariant atomistic force fields (MACE-MH-1 \cite{batatia2022mace,batatia2025crosslearning}, MACE-MPA-0 \cite{batatia2025foundation}, GRACE-2L-OMAT \cite{lysogorskiy2025graph} and MatterSim (v1 5M) \cite{yang2024mattersim}) against PBE-\ac{DFT} calculations, a level of theory chosen to match the training data of these foundation models. 
We present the antimony vacancy in Sb$_2$Se$_3$ as a representative system due to its rich charge-state multiplicity ($q = -3$ to $+1$) and complex defect landscapes \cite{wang2024upper}. 
\ce{Sb2Se3} crystallizes in the \textit{Pnma} space group with two inequivalent Sb sites and three inequivalent Se sites (Fig. S1).
The foundation model benchmarks reported here were performed for $V_{\text{Sb(1)}}$.
To systematically explore these \ac{PESs}, we employed \textsc{ShakeNBreak} \cite{mosquera2022shakenbreak} to generate diverse initial bond-distorted configurations for each charge state.
\textsc{ShakeNBreak} determines the number of nearest-neighbor bonds to distort according to the defect's effective excess charge, leading to different initial geometries for different charge states.
The same initial configurations were used for both foundation model and \ac{DFT} relaxations in each charge state to ensure a fair comparison.
For the fully ionized state ($q = -3$), where the effective excess charge is zero, the initial structures from the neutral state were used to ensure a comprehensive search.
These structures were subsequently relaxed using each foundation model to identify the predicted ground-state structures.

Fig. \ref{fig_1} illustrates the failure of bulk-trained foundation models when applied to charged point defects.
As shown in Fig. \ref{fig_1}(a), none of the tested models consistently identifies the \ac{DFT}-predicted ground-state geometry of $V_{\text{Sb(1)}}$ across all charge states. The structural \ac{RMSD} generally ranges from 0.2 to 0.4 Å, well beyond the 0.1 Å threshold for reliable structure identification. 
Although GRACE-2L-OMAT and MACE-MH-1 appear to match the DFT structure at $q = -2$ and $q = -1$ respectively, we show below that these are accidental agreements limited to a single charge state and resulting from error cancellation rather than physically accurate modeling.

We attribute these failures to two key discrepancies between current foundation models and defect physics:
a domain gap in the training data, and the absence of charge-state descriptors in the model architectures.
First, foundation force fields are trained predominantly on closed-shell bulk structures, so they tend to treat point defects as steric voids instead of electronically active centers that drive local reconstruction.
In the neutral state of $V_{\text{Sb(1)}}$ (Fig. \ref{fig_1}(b)), removing an Sb atom introduces three holes that drive a covalent reconstruction into a Se trimer with a bond angle of 110$^\circ$ through electron sharing.
Since these electronic environments are absent from the bulk training data, none of the tested \ac{MLIPs} reproduces this reconstruction correctly. 
Instead, they tend to relax into high-symmetry, bulk-like configurations or fall into a spurious local minimum with a bond angle near 99$^\circ$.
This shows that even neutral defects, which dominate the existing \ac{MLIP} literature, are already challenging cases for current foundation models.
Second, standard \ac{MLIPs} represent the total energy as a function of atomic positions and species without incorporating any charge information.
As a result, they assign identical local forces and energies to configurations with the same local geometry, collapsing charge-dependent \ac{PESs} into a single landscape.
Even for the fully ionized $q = -3$ state (Fig. \ref{fig_1}(c)), which corresponds to a closed-shell configuration with no localized charge and minimal structural reconstruction, the models still fail to identify the DFT reference minimum.
Although they perform slightly better in this regime with lower \ac{RMSD} values, all predictions still exceed the 0.1 Å threshold. 
This improvement arises from the models' preference for low-distortion geometries close to the bulk structure environments. Without explicit descriptors for the charge state, the models cannot distinguish the charge-induced relaxation from the pristine structure.

Given these limitations, the isolated successes of GRACE-2L-OMAT ($q=-2$) and MACE-MH-1 ($q=-1$) in Fig. \ref{fig_1}(a) do not reflect physically meaningful predictions.
For $V_{\text{Sb(1)}}$, the $q=-2$ state adopts a geometry similar to the fully ionized $q=-3$ case, while the $q=-1$ state is non-magnetic with reduced covalent reconstruction compared to the neutral state (Fig. S4).
These intermediate charge states coincidentally match foundation models' bias toward bulk-like configurations.
As shown in Fig. \ref{fig_1}(b-c) (and Fig. S3 for the other charge states), the models systematically predict flatter \ac{PESs} compared to \ac{DFT}, indicating incorrect forces.
Consequently, such coincidental structural agreements resulting from error cancellation are not transferable and cannot be relied upon for predictive defect engineering.
Furthermore, subsequent DFT relaxations initiated from foundation-model minima do not consistently converge to the DFT reference ground states across all charge states (Section S3.1.2, Fig S4). 

These results indicate that foundation models trained solely on bulk structures, even when highly accurate for ideal crystals \cite{loew2025universal}, are fundamentally unreliable for defect physics in semiconductors and insulators.
This limitation applies across all charge states: not only to charged defects, where charge-state descriptors are absent, but also to neutral defects, where excess carriers can drive reconstructions outside the training domain.
Compared to standard \ac{MLIPs} for pristine materials, reliable descriptions of charged point defects require two additional factors.
First, the training data must cover relevant configurations beyond the distribution found in pristine crystals. 
Second, the model architecture must explicitly encode electronic state information to distinguish between different electronic configurations of the same defect. 
This can be achieved through approaches such as global embeddings of electronic state variables, including charge and spin, or explicit electrostatic models \cite{ko2021fourth,cheng2025latent,fan2026qnep}.
Further benchmarks show that charge-aware foundation models that explicitly impose the total charge but lack relevant defect training data still fail to reproduce the DFT ground states (Section S3.2).

\begin{figure*}[ht]
    \centering    {\includegraphics[width=0.7\textwidth]{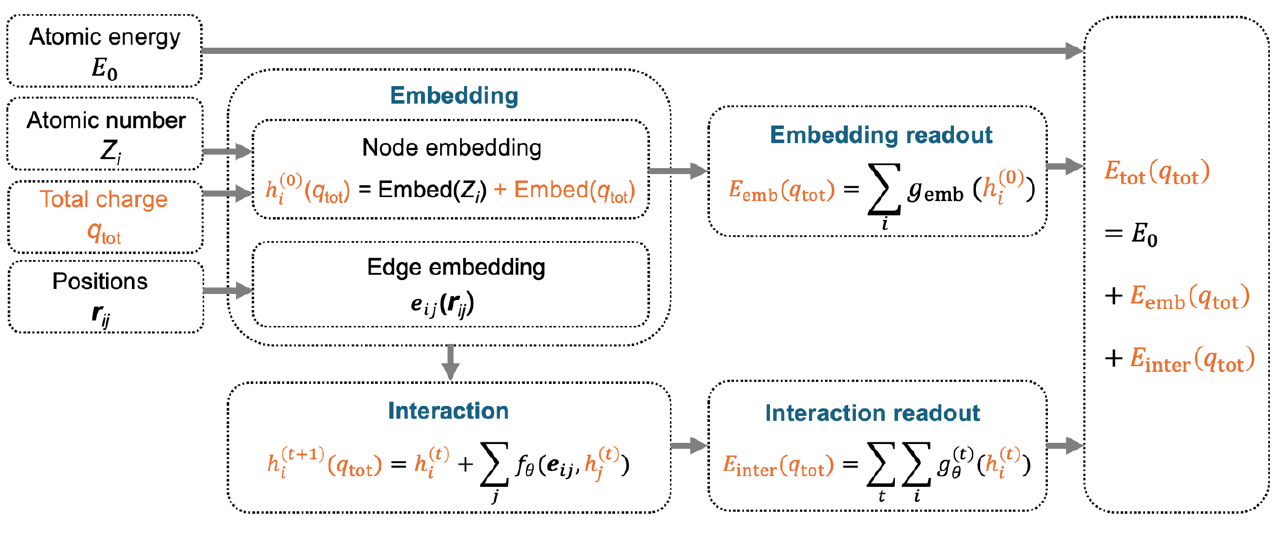}} \\
    \caption{Schematic diagram for the integration of global charge embeddings for point defects in machine learning interatomic potentials (MACE \cite{batatia2022mace} architecture). 
    The total charge $q_\textrm{tot}$ is added to each atomic species embedding during feature initialization.
    This charge conditioning propagates through the interaction layers, where node features $h_i^{(t)}$ are iteratively updated via a learned interaction function $f_\theta$ that aggregates messages from edge embeddings $\boldsymbol{e}_{ij}$ and neighboring features $h_j^{(t)}$.
    The interaction energy $E_\textrm{inter}$ is computed by summing the readout outputs $g_\theta^{(t)}(h_i^{(t)})$ over all atoms $i$ and message-passing iterations $t$, while a geometry-independent energy term $E_\textrm{emb}$ is extracted from the initial charge-enriched embeddings.
    The total energy $E_\textrm{tot}$ is defined as the sum of the atomic reference energy $E_0$, the interaction energy $E_\textrm{inter}$ and the embedding energy $E_\textrm{emb}$, with $E_0 = \sum_i E_{\textrm{ref}}(Z_i)$ denoting the sum of atomic reference energies.
    Atomic forces are computed identically to standard MACE, with \( \mathbf{F}_i = -\partial E_{\mathrm{tot}}/\partial \mathbf{r}_i = -\partial E_{\mathrm{inter}}/\partial \mathbf{r}_i \),
    since both the atomic reference energy $E_0$ and the embedding energy \(E_{\mathrm{emb}}\) are independent of atomic positions.
    }
    \label{fig_2}
\end{figure*}

\textbf{Charged defect prediction with global embeddings} — 
Building on this insight, we trained a bespoke joint model that satisfies both conditions.
The training dataset includes defect configurations across multiple charge states simulated by \ac{DFT}, and the architecture combines a global embedding of the total defect charge with atomic features.
The modified architecture (Fig. \ref{fig_2}) implements MACE v0.3.15 \cite{batatia2022mace}, with charge-dependent components highlighted in orange.
MACE was chosen for its many-body equivariant message passing architecture with high-order geometric features.
The total charge $q_\textrm{tot}$ is mapped to a learnable vector that is added to the atomic species embeddings.
The charge information is learned through two pathways:
(1) it propagates through the message-passing layers, enabling the network to learn charge-dependent interaction energies $E_\textrm{inter}$ that govern the potential energy surfaces and, through their ingredients, the forces driving defect reconstruction,
and (2) the charge-enriched embedding is read out directly prior to message passing to provide a geometry-independent energy contribution $E_\textrm{emb}$, capturing the charge-state energy shifts that are not represented by local atomic geometry. 

\begin{figure*}[ht]
    \centering    {\includegraphics[width=1.0\textwidth]{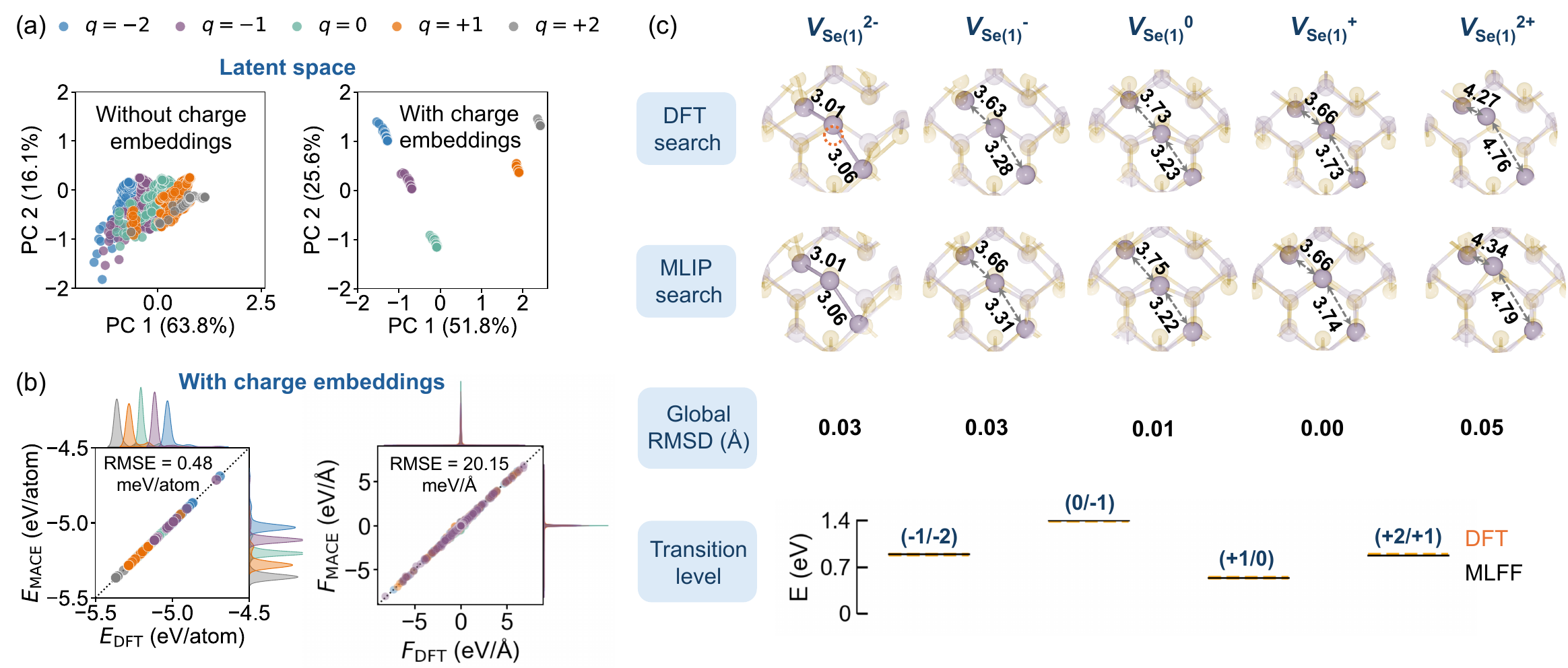}} \\
    \caption{(a) Principal component analysis (PCA) visualization of the atomic descriptor space for $V_{\text{Se(1)}}$ in \ce{Sb2Se3} across five charge states without (left) and with (right) global charge embeddings. 
    The latent space is constructed using both training and test configurations.
    Percentages in parentheses indicate the explained variance for each component.
    (b) Parity plots of total energies and atomic forces predicted by the charge-embedded MACE model compared with HSE06 DFT on the test dataset.
    Root mean square errors (RMSEs) are shown as averages across five charge states.
    (c) Comparison between the charge-embedded MACE model and HSE06 DFT for ground-state defect geometries (top) and thermodynamic transition levels (bottom). 
    For each charge state, the global minimum was identified by relaxing a set of bond-distorted structures. Key bond lengths are labeled in Å. Thermodynamic transition levels obtained from HSE06 DFT are shown as orange dashed lines, while MLIP predictions are shown as black solid lines.
    }
    \label{fig_3}
\end{figure*}

While the benchmarks in Fig. \ref{fig_1} used the PBE functional to match the training domain of foundation models, accurate defect physics often requires a higher level of theory.
Semi-local functionals like PBE systematically underestimate band gap of insulators and suffer from delocalization errors that impact defect charge distribution and energetics \cite{mori2008localization,squires2025guidelines}.
We therefore train the model using hybrid HSE06 DFT data, providing a robust reference for defect predictions.
The model is trained on raw \ac{DFT} total energies and forces without finite-size electrostatic corrections. These corrections are determined by the charge state and supercell geometry, so they do not affect atomic forces or the topology of the \ac{PES}. 
The resulting model therefore reproduces the defect landscape at the HSE06 reference level, with accuracy bounded by the underlying calculations.

While $V_{\text{Sb(1)}}$ provides a clear demonstration of foundation model failure due to its pronounced covalent reconstruction, we demonstrate the charge-embedded capabilities using $V_{\text{Se(1)}}$ in \ce{Sb2Se3} as an example due to its challenging \ac{PES} (Fig. S8). For the neutral charge state, DFT searches using \textsc{ShakeNBreak} \cite{mosquera2022shakenbreak} reveal that only 2 out of 14 bond distortions successfully identify the global minimum. Furthermore, the energy difference between the stable and metastable configurations is as small as $\sim$ 0.1 eV for a 59-atom supercell (1-2 meV/atom).
We generated training data by relaxing bond-distorted configurations across five charge states, incorporating full relaxation trajectories to sample the \ac{PES} comprehensively (see Methods in Section S2).

To visualize the role of global charge embeddings, we examined the model's latent representation of the defect.
Fig. \ref{fig_3}(a) maps the descriptor space as the average difference between the MACE feature vectors of each defect configuration and the pristine reference.
This metric captures the collective local environment perturbations induced by the vacancy.
Without global charge embeddings (Fig. \ref{fig_3}(a), left), configurations from different charge states overlap significantly in the \ac{2D} descriptor space, even though the first two principal components capture 79.5 $\%$ of the variance.
This highlights a fundamental limitation of \ac{MLIPs} without charge attributes, which rely solely on elemental and structural differences to distinguish between charge states. 

Specifically, in bulk-like regions far from the defect site, local environments remain nearly identical across charge states, while total energies differ due to different electronic occupation. 
This leads to a non-unique mapping between local descriptors and energy labels.
In contrast, with charge embeddings included (right panel), the descriptors segregate into distinct, charge-specific clusters while retaining sensitivity to local structural variations.
This separation allows the model to learn charge-dependent energy offsets for structurally similar configurations, enabling accurate predictions across all five charge states (with averaged RMSE$_E$ = 0.48 meV/atom and RMSE$_F$ = 20.15 meV/Å, Fig. \ref{fig_3}(b)).
As shown in Fig. \ref{fig_3}(c), the charge-embedded model successfully identifies the HSE06 reference ground states for all cases, with global \ac{RMSD} values within 0.05 Å and well below the 0.1 Å threshold for reliable structure identification. 
The model reproduces the local vacancy environments with negligible deviation, capturing the subtle charge-dependent reconstructions that foundation models fail to describe.
Beyond ground-state structures, the model provides a quantitatively accurate description of the energetics of the defects.
It correctly distinguishes meta-stable configurations with energy differences as small as 1 meV/atom (Fig. S8), and predicts thermodynamic transition levels (Fig. \ref{fig_3}, bottom) that closely match the HSE06 reference across all charge transitions, with a maximum deviation of only 0.015 eV (Table S3).
These results confirm that combining explicit charge embeddings with targeted defect training data enables a joint model to achieve ab initio accuracy across multiple charge states, even for highly complex defect landscapes.
Additional benchmarks show that explicitly modeling atomic charge distribution or electrostatics with QET \cite{ko2025fast}, MACE-POLAR \cite{batatia2026mace} and MACE-LES \cite{cheng2025latent} gives a less accurate description of the DFT defect \ac{PESs} (Section S5).

\begin{figure*}[ht]
    \centering    {\includegraphics[width=0.8\textwidth]{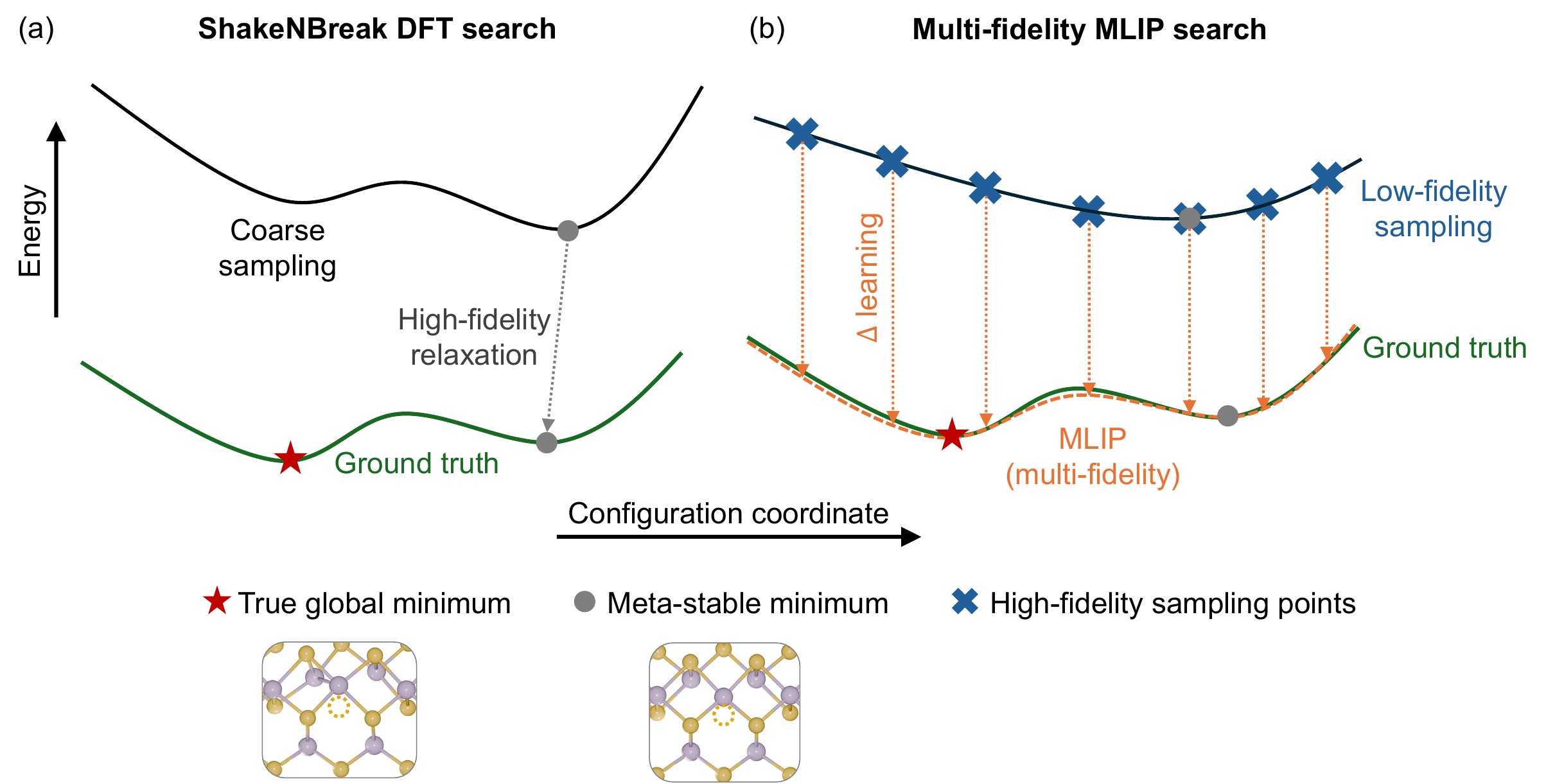}} \\
    \caption{Schematic diagram of structure searching for neutral $V_\textrm{Se(1)}$ using
    (a) a standard workflow with \textsc{ShakeNBreak} \cite{mosquera2022shakenbreak} and 
    (b) a multi-fidelity (MF) strategy with a MLIP. 
    The green curves correspond to high-fidelity (HF) \ac{PESs} calculated using HSE06/DFT with converged \textit{k}-point sampling. 
    The red star and grey circle denote the global minimum and a meta-stable minimum on this reference \ac{PES}, respectively.
    (a) The black curve represents the coarse \ac{PES} sampling using HSE06 with $\Gamma$-point sampling, and the grey dashed arrow indicates the \ac{HF} relaxation starting from the coarse sampling minimum.
    (b) The blue curve represents the low-fidelity \ac{PES} obtained from PBE with converged \textit{k}-point sampling. 
    Blue crosses mark \ac{HF} single-point calculations performed on a subset of the PBE-relaxed structures.
    Orange vertical arrows illustrate the $\Delta$-learning corrections learnt by the \ac{MF} model.
    }
    \label{fig_4}
\end{figure*}

\textbf{Improved accuracy and efficiency via multi-fidelity training} — 
While the charge-embedded model discussed above achieves accurate defect predictions, its reliance on full relaxation trajectories at the hybrid functional level presents a significant bottleneck for large-scale applications. 
To reduce the cost of expensive hybrid functional relaxations, commonly used workflows like \textsc{ShakeNBreak} \cite{mosquera2022shakenbreak} use a two-stage strategy \cite{mosquera2023identifying}. 
The \ac{PES} is initially explored using lower-cost methods (e.g., $\Gamma$-point sampling) to identify the minima, followed by higher-fidelity calculations only for the identified minima (Fig. \ref{fig_4}(a)). 
However, this approach assumes that the coarse-level search adequately reproduces the shape of \ac{HF} \ac{PESs} and its global minimum.
Furthermore, even when limited to $\Gamma$-point sampling, the computational cost remains substantial for complex defect landscapes (Section S6.3).

Motivated by these limitations, we developed a \ac{MF} training strategy (Fig. \ref{fig_4}(b)) that samples the \ac{PES} densely at the low-cost PBE level while incorporating a sparse subset ($\sim$10\%) of \ac{HF} HSE06 corrections determined using stratified sampling.
The model learns the functional-dependent difference ($\Delta$) from HSE06 single-point energies on selected PBE structures through a joint training framework with a shared latent representation and separate (low fidelity) PBE and (high fidelity) HSE06 readout channels.
Using neutral $V_{\text{Se(1)}}$ as an example, the \ac{MF} approach reveals the limitation of standard defect searching workflows.
As shown in Fig. \ref{fig_4}, conventional coarse search using HSE06 at the $\Gamma$-point (Fig. S15) identifies only a meta-stable minimum (gray circle).
In contrast, the \ac{MF}-trained model discovers a distinct global minimum 0.02 eV lower in energy (red star) by learning the functional-dependent correction required to stabilize this basin (Fig. \ref{fig_4}(b)).
This prediction is further confirmed by hybrid-functional DFT with converged \textit{k}-point sampling, which reproduces the ML-predicted minimum within 8 meV.
Beyond improved accuracy, this \ac{MF} approach dramatically reduces the computational cost of comprehensive PES exploration by approximately three orders of magnitude (Section S6.3), enabling systematic studies across multiple charge states and defect types.

Moreover, once trained, the \ac{MF} model serves as an efficient surrogate for the hybrid-functional \ac{PES}, enabling access to properties that require dense sampling of the defect landscapes that are otherwise prohibitively expensive at the hybrid-functional level.
We demonstrate this by calculating \ac{CCDs}, which describe charge carrier capture processes that limit the power conversion efficiency in optoelectronic devices \cite{alkauskas2014first}.
Constructing a \ac{CCD} requires evaluating the energies of two charge states along a continuous reaction coordinate connecting their ground-state geometries, with sufficient sampling to capture their curvature and the crossing region.
Our \ac{MF} model reproduces HSE06 reference CCDs with high precision at negligible inference cost (Fig. S17).
This approach can be extended to other thermodynamic and spectroscopic properties, such as finite-temperature free energies \cite{mosquera2025point,hainer2025thermalstabilizationdefectcharge}, luminescence lineshapes \cite{turiansky2025machinelearningphononspectra,Linderlv2025} and defect dynamics \cite{Tyagi2025}, offering new possibilities for accelerated defect screening and property prediction with hybrid-level accuracy.

In conclusion, as machine learning force fields have matured in their description of bulk materials, we evaluate their generalizability to the out-of-distribution regime of charged point defects.
Using vacancies in \ce{Sb2Se3} as a case study, we demonstrate that while current foundation models fail to capture the defect energy landscapes, training on defect configurations with global charge embeddings at a fixed defect supercell size can recover ground-state structures to within 0.05 Å, with defect formation energies and thermodynamic transition levels reproduced at hybrid-level DFT accuracy to within 0.02 eV.
We further introduce a multi-fidelity training scheme that not only discovers global minima missed by standard defect screening workflows, but also provides a computationally efficient surrogate model that can be used to access larger configurational spaces. 
Together, these strategies offer a practical framework for bridging the gap between \textit{ab initio} accuracy and large-scale defect modeling, enabling high-throughput prediction of charged defect structures, energetics, and related physical properties.
More broadly, our work lays the groundwork for future defect foundation models that generalize across materials, charge states, and levels of theory.
Broad coverage of defect environments, learnable charge conditioning and multi-fidelity training together provide a practical basis for extending pristine-material foundation models to charged-defect applications.


\section{Data Availability Statement}
The datasets generated and analysed during the current study will be made available upon publication. This includes the DFT reference data and the curated training/test splits for charged defects in Sb$_{2}$Se$_{3}$. The trained MACE model weights, along with the inference scripts required to reproduce the structural searches and thermodynamic transition level predictions, are also deposited at the same location. The MACE software package is open-source and available at \url{https://github.com/ACEsuit/mace}.

\section{Acknowledgments}
We thank Alex M. Ganose for helpful discussions and support.
This research received support through Schmidt Sciences, LLC.
I.M.-L. thanks Imperial College London (ICL) for funding a President’s PhD scholarship. 
Via our membership of the UK's HEC Materials Chemistry Consortium, which is funded by EPSRC (EP/X035859/1), this work used the ARCHER2 UK National Supercomputing Service (http://www.archer2.ac.uk).
We are grateful to the UK Materials and Molecular Modelling Hub for computational resources, which is partially funded by EPSRC (EP/T022213/1, EP/W032260/1 and EP/P020194/1).

\bibliographystyle{apsrev4-2}
\bibliography{ref}

\begin{thebibliography}{50}%
\makeatletter
\providecommand \@ifxundefined [1]{%
 \@ifx{#1\undefined}
}%
\providecommand \@ifnum [1]{%
 \ifnum #1\expandafter \@firstoftwo
 \else \expandafter \@secondoftwo
 \fi
}%
\providecommand \@ifx [1]{%
 \ifx #1\expandafter \@firstoftwo
 \else \expandafter \@secondoftwo
 \fi
}%
\providecommand \natexlab [1]{#1}%
\providecommand \enquote  [1]{``#1''}%
\providecommand \bibnamefont  [1]{#1}%
\providecommand \bibfnamefont [1]{#1}%
\providecommand \citenamefont [1]{#1}%
\providecommand \href@noop [0]{\@secondoftwo}%
\providecommand \href [0]{\begingroup \@sanitize@url \@href}%
\providecommand \@href[1]{\@@startlink{#1}\@@href}%
\providecommand \@@href[1]{\endgroup#1\@@endlink}%
\providecommand \@sanitize@url [0]{\catcode `\\12\catcode `\$12\catcode `\&12\catcode `\#12\catcode `\^12\catcode `\_12\catcode `\%12\relax}%
\providecommand \@@startlink[1]{}%
\providecommand \@@endlink[0]{}%
\providecommand \url  [0]{\begingroup\@sanitize@url \@url }%
\providecommand \@url [1]{\endgroup\@href {#1}{\urlprefix }}%
\providecommand \urlprefix  [0]{URL }%
\providecommand \Eprint [0]{\href }%
\providecommand \doibase [0]{https://doi.org/}%
\providecommand \selectlanguage [0]{\@gobble}%
\providecommand \bibinfo  [0]{\@secondoftwo}%
\providecommand \bibfield  [0]{\@secondoftwo}%
\providecommand \translation [1]{[#1]}%
\providecommand \BibitemOpen [0]{}%
\providecommand \bibitemStop [0]{}%
\providecommand \bibitemNoStop [0]{.\EOS\space}%
\providecommand \EOS [0]{\spacefactor3000\relax}%
\providecommand \BibitemShut  [1]{\csname bibitem#1\endcsname}%
\let\auto@bib@innerbib\@empty
\bibitem [{\citenamefont {Freysoldt}\ \emph {et~al.}(2014)\citenamefont {Freysoldt}, \citenamefont {Grabowski}, \citenamefont {Hickel}, \citenamefont {Neugebauer}, \citenamefont {Kresse}, \citenamefont {Janotti},\ and\ \citenamefont {Van~de Walle}}]{freysoldt2014first}%
  \BibitemOpen
  \bibfield  {author} {\bibinfo {author} {\bibfnamefont {C.}~\bibnamefont {Freysoldt}}, \bibinfo {author} {\bibfnamefont {B.}~\bibnamefont {Grabowski}}, \bibinfo {author} {\bibfnamefont {T.}~\bibnamefont {Hickel}}, \bibinfo {author} {\bibfnamefont {J.}~\bibnamefont {Neugebauer}}, \bibinfo {author} {\bibfnamefont {G.}~\bibnamefont {Kresse}}, \bibinfo {author} {\bibfnamefont {A.}~\bibnamefont {Janotti}},\ and\ \bibinfo {author} {\bibfnamefont {C.~G.}\ \bibnamefont {Van~de Walle}},\ }\href@noop {} {\bibfield  {journal} {\bibinfo  {journal} {Rev. Mod. Phys.}\ }\textbf {\bibinfo {volume} {86}},\ \bibinfo {pages} {253} (\bibinfo {year} {2014})}\BibitemShut {NoStop}%
\bibitem [{\citenamefont {Walsh}\ and\ \citenamefont {Zunger}(2017)}]{walsh2017instilling}%
  \BibitemOpen
  \bibfield  {author} {\bibinfo {author} {\bibfnamefont {A.}~\bibnamefont {Walsh}}\ and\ \bibinfo {author} {\bibfnamefont {A.}~\bibnamefont {Zunger}},\ }\href@noop {} {\bibfield  {journal} {\bibinfo  {journal} {Nat. Mater.}\ }\textbf {\bibinfo {volume} {16}},\ \bibinfo {pages} {964} (\bibinfo {year} {2017})}\BibitemShut {NoStop}%
\bibitem [{\citenamefont {Mosquera-Lois}\ and\ \citenamefont {Kavanagh}(2021)}]{mosquera-lois_search_2021}%
  \BibitemOpen
  \bibfield  {author} {\bibinfo {author} {\bibfnamefont {I.}~\bibnamefont {Mosquera-Lois}}\ and\ \bibinfo {author} {\bibfnamefont {S.~R.}\ \bibnamefont {Kavanagh}},\ }\href {https://doi.org/10.1016/j.matt.2021.06.003} {\bibfield  {journal} {\bibinfo  {journal} {Matter}\ }\textbf {\bibinfo {volume} {4}},\ \bibinfo {pages} {2602} (\bibinfo {year} {2021})}\BibitemShut {NoStop}%
\bibitem [{\citenamefont {Kavanagh}\ \emph {et~al.}(2022)\citenamefont {Kavanagh}, \citenamefont {Scanlon}, \citenamefont {Walsh},\ and\ \citenamefont {Freysoldt}}]{kavanagh2022impact}%
  \BibitemOpen
  \bibfield  {author} {\bibinfo {author} {\bibfnamefont {S.~R.}\ \bibnamefont {Kavanagh}}, \bibinfo {author} {\bibfnamefont {D.~O.}\ \bibnamefont {Scanlon}}, \bibinfo {author} {\bibfnamefont {A.}~\bibnamefont {Walsh}},\ and\ \bibinfo {author} {\bibfnamefont {C.}~\bibnamefont {Freysoldt}},\ }\href@noop {} {\bibfield  {journal} {\bibinfo  {journal} {Faraday Discuss.}\ }\textbf {\bibinfo {volume} {239}},\ \bibinfo {pages} {339} (\bibinfo {year} {2022})}\BibitemShut {NoStop}%
\bibitem [{\citenamefont {Wang}\ \emph {et~al.}(2023)\citenamefont {Wang}, \citenamefont {Kavanagh}, \citenamefont {Scanlon},\ and\ \citenamefont {Walsh}}]{wang2023fourelectron}%
  \BibitemOpen
  \bibfield  {author} {\bibinfo {author} {\bibfnamefont {X.}~\bibnamefont {Wang}}, \bibinfo {author} {\bibfnamefont {S.~R.}\ \bibnamefont {Kavanagh}}, \bibinfo {author} {\bibfnamefont {D.~O.}\ \bibnamefont {Scanlon}},\ and\ \bibinfo {author} {\bibfnamefont {A.}~\bibnamefont {Walsh}},\ }\href@noop {} {\bibfield  {journal} {\bibinfo  {journal} {Phys. Rev. B}\ }\textbf {\bibinfo {volume} {108}},\ \bibinfo {pages} {134102} (\bibinfo {year} {2023})}\BibitemShut {NoStop}%
\bibitem [{\citenamefont {El-Mellouhi}\ \emph {et~al.}(2004)\citenamefont {El-Mellouhi}, \citenamefont {Mousseau},\ and\ \citenamefont {Ordej{\'o}n}}]{el2004sampling}%
  \BibitemOpen
  \bibfield  {author} {\bibinfo {author} {\bibfnamefont {F.}~\bibnamefont {El-Mellouhi}}, \bibinfo {author} {\bibfnamefont {N.}~\bibnamefont {Mousseau}},\ and\ \bibinfo {author} {\bibfnamefont {P.}~\bibnamefont {Ordej{\'o}n}},\ }\href@noop {} {\bibfield  {journal} {\bibinfo  {journal} {Phys. Rev. B}\ }\textbf {\bibinfo {volume} {70}},\ \bibinfo {pages} {205202} (\bibinfo {year} {2004})}\BibitemShut {NoStop}%
\bibitem [{\citenamefont {Squires}\ \emph {et~al.}(2024)\citenamefont {Squires}, \citenamefont {Ganeshkumar}, \citenamefont {Savory}, \citenamefont {Kavanagh},\ and\ \citenamefont {Scanlon}}]{Squires2024}%
  \BibitemOpen
  \bibfield  {author} {\bibinfo {author} {\bibfnamefont {A.~G.}\ \bibnamefont {Squires}}, \bibinfo {author} {\bibfnamefont {L.}~\bibnamefont {Ganeshkumar}}, \bibinfo {author} {\bibfnamefont {C.~N.}\ \bibnamefont {Savory}}, \bibinfo {author} {\bibfnamefont {S.~R.}\ \bibnamefont {Kavanagh}},\ and\ \bibinfo {author} {\bibfnamefont {D.~O.}\ \bibnamefont {Scanlon}},\ }\href {https://doi.org/10.1021/acsenergylett.4c01307} {\bibfield  {journal} {\bibinfo  {journal} {ACS Energy Lett.}\ }\textbf {\bibinfo {volume} {9}},\ \bibinfo {pages} {4180–4187} (\bibinfo {year} {2024})}\BibitemShut {NoStop}%
\bibitem [{\citenamefont {Mosquera-Lois}\ \emph {et~al.}(2022)\citenamefont {Mosquera-Lois}, \citenamefont {Kavanagh}, \citenamefont {Walsh},\ and\ \citenamefont {Scanlon}}]{mosquera2022shakenbreak}%
  \BibitemOpen
  \bibfield  {author} {\bibinfo {author} {\bibfnamefont {I.}~\bibnamefont {Mosquera-Lois}}, \bibinfo {author} {\bibfnamefont {S.~R.}\ \bibnamefont {Kavanagh}}, \bibinfo {author} {\bibfnamefont {A.}~\bibnamefont {Walsh}},\ and\ \bibinfo {author} {\bibfnamefont {D.~O.}\ \bibnamefont {Scanlon}},\ }\href@noop {} {\bibfield  {journal} {\bibinfo  {journal} {J. Open Source Softw.}\ }\textbf {\bibinfo {volume} {7}},\ \bibinfo {pages} {4817} (\bibinfo {year} {2022})}\BibitemShut {NoStop}%
\bibitem [{\citenamefont {Pickard}\ and\ \citenamefont {Needs}(2011)}]{pickard2011ab}%
  \BibitemOpen
  \bibfield  {author} {\bibinfo {author} {\bibfnamefont {C.~J.}\ \bibnamefont {Pickard}}\ and\ \bibinfo {author} {\bibfnamefont {R.}~\bibnamefont {Needs}},\ }\href@noop {} {\bibfield  {journal} {\bibinfo  {journal} {J. Phys.: Condens. Matter}\ }\textbf {\bibinfo {volume} {23}},\ \bibinfo {pages} {053201} (\bibinfo {year} {2011})}\BibitemShut {NoStop}%
\bibitem [{\citenamefont {Wang}\ \emph {et~al.}(2024{\natexlab{a}})\citenamefont {Wang}, \citenamefont {Kavanagh},\ and\ \citenamefont {Walsh}}]{wang2024sulfur}%
  \BibitemOpen
  \bibfield  {author} {\bibinfo {author} {\bibfnamefont {X.}~\bibnamefont {Wang}}, \bibinfo {author} {\bibfnamefont {S.~R.}\ \bibnamefont {Kavanagh}},\ and\ \bibinfo {author} {\bibfnamefont {A.}~\bibnamefont {Walsh}},\ }\href@noop {} {\bibfield  {journal} {\bibinfo  {journal} {ACS Energy Lett.}\ }\textbf {\bibinfo {volume} {10}},\ \bibinfo {pages} {161} (\bibinfo {year} {2024}{\natexlab{a}})}\BibitemShut {NoStop}%
\bibitem [{\citenamefont {Batatia}\ \emph {et~al.}(2025{\natexlab{a}})\citenamefont {Batatia}, \citenamefont {Lin}, \citenamefont {Hart}, \citenamefont {Kasoar}, \citenamefont {Elena}, \citenamefont {Norwood}, \citenamefont {Wolf},\ and\ \citenamefont {Cs{\'a}nyi}}]{batatia2025crosslearning}%
  \BibitemOpen
  \bibfield  {author} {\bibinfo {author} {\bibfnamefont {I.}~\bibnamefont {Batatia}}, \bibinfo {author} {\bibfnamefont {C.}~\bibnamefont {Lin}}, \bibinfo {author} {\bibfnamefont {J.}~\bibnamefont {Hart}}, \bibinfo {author} {\bibfnamefont {E.}~\bibnamefont {Kasoar}}, \bibinfo {author} {\bibfnamefont {A.~M.}\ \bibnamefont {Elena}}, \bibinfo {author} {\bibfnamefont {S.~W.}\ \bibnamefont {Norwood}}, \bibinfo {author} {\bibfnamefont {T.}~\bibnamefont {Wolf}},\ and\ \bibinfo {author} {\bibfnamefont {G.}~\bibnamefont {Cs{\'a}nyi}},\ }\href@noop {} {\bibfield  {journal} {\bibinfo  {journal} {arXiv}\ } (\bibinfo {year} {2025}{\natexlab{a}})},\ \bibinfo {note} {arXiv:2510.25380}\BibitemShut {NoStop}%
\bibitem [{\citenamefont {Batatia}\ \emph {et~al.}(2025{\natexlab{b}})\citenamefont {Batatia}, \citenamefont {Benner}, \citenamefont {Chiang}, \citenamefont {Elena}, \citenamefont {Kov{\'a}cs}, \citenamefont {Riebesell}, \citenamefont {Advincula}, \citenamefont {Asta}, \citenamefont {Avaylon}, \citenamefont {Baldwin} \emph {et~al.}}]{batatia2025foundation}%
  \BibitemOpen
  \bibfield  {author} {\bibinfo {author} {\bibfnamefont {I.}~\bibnamefont {Batatia}}, \bibinfo {author} {\bibfnamefont {P.}~\bibnamefont {Benner}}, \bibinfo {author} {\bibfnamefont {Y.}~\bibnamefont {Chiang}}, \bibinfo {author} {\bibfnamefont {A.~M.}\ \bibnamefont {Elena}}, \bibinfo {author} {\bibfnamefont {D.~P.}\ \bibnamefont {Kov{\'a}cs}}, \bibinfo {author} {\bibfnamefont {J.}~\bibnamefont {Riebesell}}, \bibinfo {author} {\bibfnamefont {X.~R.}\ \bibnamefont {Advincula}}, \bibinfo {author} {\bibfnamefont {M.}~\bibnamefont {Asta}}, \bibinfo {author} {\bibfnamefont {M.}~\bibnamefont {Avaylon}}, \bibinfo {author} {\bibfnamefont {W.~J.}\ \bibnamefont {Baldwin}}, \emph {et~al.},\ }\href@noop {} {\bibfield  {journal} {\bibinfo  {journal} {J. Chem. Phys.}\ }\textbf {\bibinfo {volume} {163}} (\bibinfo {year} {2025}{\natexlab{b}})}\BibitemShut {NoStop}%
\bibitem [{\citenamefont {Lysogorskiy}\ \emph {et~al.}(2025)\citenamefont {Lysogorskiy}, \citenamefont {Bochkarev},\ and\ \citenamefont {Drautz}}]{lysogorskiy2025graph}%
  \BibitemOpen
  \bibfield  {author} {\bibinfo {author} {\bibfnamefont {Y.}~\bibnamefont {Lysogorskiy}}, \bibinfo {author} {\bibfnamefont {A.}~\bibnamefont {Bochkarev}},\ and\ \bibinfo {author} {\bibfnamefont {R.}~\bibnamefont {Drautz}},\ }\href@noop {} {\bibfield  {journal} {\bibinfo  {journal} {arXiv}\ } (\bibinfo {year} {2025})},\ \bibinfo {note} {arXiv:2508.17936}\BibitemShut {NoStop}%
\bibitem [{\citenamefont {Yang}\ \emph {et~al.}(2024)\citenamefont {Yang}, \citenamefont {Hu}, \citenamefont {Zhou}, \citenamefont {Liu}, \citenamefont {Shi}, \citenamefont {Li}, \citenamefont {Li}, \citenamefont {Chen}, \citenamefont {Chen}, \citenamefont {Zeni} \emph {et~al.}}]{yang2024mattersim}%
  \BibitemOpen
  \bibfield  {author} {\bibinfo {author} {\bibfnamefont {H.}~\bibnamefont {Yang}}, \bibinfo {author} {\bibfnamefont {C.}~\bibnamefont {Hu}}, \bibinfo {author} {\bibfnamefont {Y.}~\bibnamefont {Zhou}}, \bibinfo {author} {\bibfnamefont {X.}~\bibnamefont {Liu}}, \bibinfo {author} {\bibfnamefont {Y.}~\bibnamefont {Shi}}, \bibinfo {author} {\bibfnamefont {J.}~\bibnamefont {Li}}, \bibinfo {author} {\bibfnamefont {G.}~\bibnamefont {Li}}, \bibinfo {author} {\bibfnamefont {Z.}~\bibnamefont {Chen}}, \bibinfo {author} {\bibfnamefont {S.}~\bibnamefont {Chen}}, \bibinfo {author} {\bibfnamefont {C.}~\bibnamefont {Zeni}}, \emph {et~al.},\ }\href@noop {} {\bibfield  {journal} {\bibinfo  {journal} {arXiv}\ } (\bibinfo {year} {2024})},\ \bibinfo {note} {arXiv:2405.04967}\BibitemShut {NoStop}%
\bibitem [{\citenamefont {Jain}\ \emph {et~al.}(2013)\citenamefont {Jain}, \citenamefont {Ong}, \citenamefont {Hautier}, \citenamefont {Chen}, \citenamefont {Richards}, \citenamefont {Dacek}, \citenamefont {Cholia}, \citenamefont {Gunter}, \citenamefont {Skinner}, \citenamefont {Ceder} \emph {et~al.}}]{jain2013commentary}%
  \BibitemOpen
  \bibfield  {author} {\bibinfo {author} {\bibfnamefont {A.}~\bibnamefont {Jain}}, \bibinfo {author} {\bibfnamefont {S.~P.}\ \bibnamefont {Ong}}, \bibinfo {author} {\bibfnamefont {G.}~\bibnamefont {Hautier}}, \bibinfo {author} {\bibfnamefont {W.}~\bibnamefont {Chen}}, \bibinfo {author} {\bibfnamefont {W.~D.}\ \bibnamefont {Richards}}, \bibinfo {author} {\bibfnamefont {S.}~\bibnamefont {Dacek}}, \bibinfo {author} {\bibfnamefont {S.}~\bibnamefont {Cholia}}, \bibinfo {author} {\bibfnamefont {D.}~\bibnamefont {Gunter}}, \bibinfo {author} {\bibfnamefont {D.}~\bibnamefont {Skinner}}, \bibinfo {author} {\bibfnamefont {G.}~\bibnamefont {Ceder}}, \emph {et~al.},\ }\href@noop {} {\bibfield  {journal} {\bibinfo  {journal} {APL Mater.}\ }\textbf {\bibinfo {volume} {1}} (\bibinfo {year} {2013})}\BibitemShut {NoStop}%
\bibitem [{\citenamefont {Barroso-Luque}\ \emph {et~al.}(2024)\citenamefont {Barroso-Luque}, \citenamefont {Shuaibi}, \citenamefont {Fu}, \citenamefont {Wood}, \citenamefont {Dzamba}, \citenamefont {Gao}, \citenamefont {Rizvi}, \citenamefont {Zitnick},\ and\ \citenamefont {Ulissi}}]{barroso_omat24}%
  \BibitemOpen
  \bibfield  {author} {\bibinfo {author} {\bibfnamefont {L.}~\bibnamefont {Barroso-Luque}}, \bibinfo {author} {\bibfnamefont {M.}~\bibnamefont {Shuaibi}}, \bibinfo {author} {\bibfnamefont {X.}~\bibnamefont {Fu}}, \bibinfo {author} {\bibfnamefont {B.~M.}\ \bibnamefont {Wood}}, \bibinfo {author} {\bibfnamefont {M.}~\bibnamefont {Dzamba}}, \bibinfo {author} {\bibfnamefont {M.}~\bibnamefont {Gao}}, \bibinfo {author} {\bibfnamefont {A.}~\bibnamefont {Rizvi}}, \bibinfo {author} {\bibfnamefont {C.~L.}\ \bibnamefont {Zitnick}},\ and\ \bibinfo {author} {\bibfnamefont {Z.~W.}\ \bibnamefont {Ulissi}},\ }\href@noop {} {\bibfield  {journal} {\bibinfo  {journal} {arXiv}\ } (\bibinfo {year} {2024})},\ \bibinfo {note} {arXiv:2410.12771}\BibitemShut {NoStop}%
\bibitem [{\citenamefont {Schmidt}\ \emph {et~al.}(2024)\citenamefont {Schmidt}, \citenamefont {Cerqueira}, \citenamefont {Romero}, \citenamefont {Loew}, \citenamefont {J{\"a}ger}, \citenamefont {Wang}, \citenamefont {Botti},\ and\ \citenamefont {Marques}}]{schmidt2024improving}%
  \BibitemOpen
  \bibfield  {author} {\bibinfo {author} {\bibfnamefont {J.}~\bibnamefont {Schmidt}}, \bibinfo {author} {\bibfnamefont {T.~F.}\ \bibnamefont {Cerqueira}}, \bibinfo {author} {\bibfnamefont {A.~H.}\ \bibnamefont {Romero}}, \bibinfo {author} {\bibfnamefont {A.}~\bibnamefont {Loew}}, \bibinfo {author} {\bibfnamefont {F.}~\bibnamefont {J{\"a}ger}}, \bibinfo {author} {\bibfnamefont {H.-C.}\ \bibnamefont {Wang}}, \bibinfo {author} {\bibfnamefont {S.}~\bibnamefont {Botti}},\ and\ \bibinfo {author} {\bibfnamefont {M.~A.}\ \bibnamefont {Marques}},\ }\href@noop {} {\bibfield  {journal} {\bibinfo  {journal} {Mater. Today Phys.}\ }\textbf {\bibinfo {volume} {48}},\ \bibinfo {pages} {101560} (\bibinfo {year} {2024})}\BibitemShut {NoStop}%
\bibitem [{\citenamefont {Loew}\ \emph {et~al.}(2025)\citenamefont {Loew}, \citenamefont {Sun}, \citenamefont {Wang}, \citenamefont {Botti},\ and\ \citenamefont {Marques}}]{loew2025universal}%
  \BibitemOpen
  \bibfield  {author} {\bibinfo {author} {\bibfnamefont {A.}~\bibnamefont {Loew}}, \bibinfo {author} {\bibfnamefont {D.}~\bibnamefont {Sun}}, \bibinfo {author} {\bibfnamefont {H.-C.}\ \bibnamefont {Wang}}, \bibinfo {author} {\bibfnamefont {S.}~\bibnamefont {Botti}},\ and\ \bibinfo {author} {\bibfnamefont {M.~A.}\ \bibnamefont {Marques}},\ }\href@noop {} {\bibfield  {journal} {\bibinfo  {journal} {npj Comput. Mater.}\ }\textbf {\bibinfo {volume} {11}},\ \bibinfo {pages} {178} (\bibinfo {year} {2025})}\BibitemShut {NoStop}%
\bibitem [{\citenamefont {Prodan}\ and\ \citenamefont {Kohn}(2005)}]{prodan2005nearsightedness}%
  \BibitemOpen
  \bibfield  {author} {\bibinfo {author} {\bibfnamefont {E.}~\bibnamefont {Prodan}}\ and\ \bibinfo {author} {\bibfnamefont {W.}~\bibnamefont {Kohn}},\ }\href@noop {} {\bibfield  {journal} {\bibinfo  {journal} {Proc. Natl. Acad. Sci. U.S.A.}\ }\textbf {\bibinfo {volume} {102}},\ \bibinfo {pages} {11635} (\bibinfo {year} {2005})}\BibitemShut {NoStop}%
\bibitem [{\citenamefont {Behler}\ and\ \citenamefont {Parrinello}(2007)}]{behler2007generalized}%
  \BibitemOpen
  \bibfield  {author} {\bibinfo {author} {\bibfnamefont {J.}~\bibnamefont {Behler}}\ and\ \bibinfo {author} {\bibfnamefont {M.}~\bibnamefont {Parrinello}},\ }\href@noop {} {\bibfield  {journal} {\bibinfo  {journal} {Phys. Rev. Lett.}\ }\textbf {\bibinfo {volume} {98}},\ \bibinfo {pages} {146401} (\bibinfo {year} {2007})}\BibitemShut {NoStop}%
\bibitem [{\citenamefont {Batatia}\ \emph {et~al.}(2022)\citenamefont {Batatia}, \citenamefont {Kovacs}, \citenamefont {Simm}, \citenamefont {Ortner},\ and\ \citenamefont {Cs{\'a}nyi}}]{batatia2022mace}%
  \BibitemOpen
  \bibfield  {author} {\bibinfo {author} {\bibfnamefont {I.}~\bibnamefont {Batatia}}, \bibinfo {author} {\bibfnamefont {D.~P.}\ \bibnamefont {Kovacs}}, \bibinfo {author} {\bibfnamefont {G.}~\bibnamefont {Simm}}, \bibinfo {author} {\bibfnamefont {C.}~\bibnamefont {Ortner}},\ and\ \bibinfo {author} {\bibfnamefont {G.}~\bibnamefont {Cs{\'a}nyi}},\ }\href@noop {} {\bibfield  {journal} {\bibinfo  {journal} {Adv. Neural Inf. Process.}\ }\textbf {\bibinfo {volume} {35}},\ \bibinfo {pages} {11423} (\bibinfo {year} {2022})}\BibitemShut {NoStop}%
\bibitem [{\citenamefont {Mannodi-Kanakkithodi}\ \emph {et~al.}(2022)\citenamefont {Mannodi-Kanakkithodi}, \citenamefont {Xiang}, \citenamefont {Jacoby}, \citenamefont {Biegaj}, \citenamefont {Dunham}, \citenamefont {Gamelin},\ and\ \citenamefont {Chan}}]{mannodi2022universal}%
  \BibitemOpen
  \bibfield  {author} {\bibinfo {author} {\bibfnamefont {A.}~\bibnamefont {Mannodi-Kanakkithodi}}, \bibinfo {author} {\bibfnamefont {X.}~\bibnamefont {Xiang}}, \bibinfo {author} {\bibfnamefont {L.}~\bibnamefont {Jacoby}}, \bibinfo {author} {\bibfnamefont {R.}~\bibnamefont {Biegaj}}, \bibinfo {author} {\bibfnamefont {S.~T.}\ \bibnamefont {Dunham}}, \bibinfo {author} {\bibfnamefont {D.~R.}\ \bibnamefont {Gamelin}},\ and\ \bibinfo {author} {\bibfnamefont {M.~K.}\ \bibnamefont {Chan}},\ }\href@noop {} {\bibfield  {journal} {\bibinfo  {journal} {Patterns}\ }\textbf {\bibinfo {volume} {3}} (\bibinfo {year} {2022})}\BibitemShut {NoStop}%
\bibitem [{\citenamefont {Mosquera-Lois}\ \emph {et~al.}(2024)\citenamefont {Mosquera-Lois}, \citenamefont {Kavanagh}, \citenamefont {Ganose},\ and\ \citenamefont {Walsh}}]{mosquera2024machine}%
  \BibitemOpen
  \bibfield  {author} {\bibinfo {author} {\bibfnamefont {I.}~\bibnamefont {Mosquera-Lois}}, \bibinfo {author} {\bibfnamefont {S.~R.}\ \bibnamefont {Kavanagh}}, \bibinfo {author} {\bibfnamefont {A.~M.}\ \bibnamefont {Ganose}},\ and\ \bibinfo {author} {\bibfnamefont {A.}~\bibnamefont {Walsh}},\ }\href@noop {} {\bibfield  {journal} {\bibinfo  {journal} {npj Comput. Mater.}\ }\textbf {\bibinfo {volume} {10}},\ \bibinfo {pages} {121} (\bibinfo {year} {2024})}\BibitemShut {NoStop}%
\bibitem [{\citenamefont {Kavanagh}(2024)}]{kavanagh2024identifying}%
  \BibitemOpen
  \bibfield  {author} {\bibinfo {author} {\bibfnamefont {S.~R.}\ \bibnamefont {Kavanagh}},\ }\href@noop {} {\bibfield  {journal} {\bibinfo  {journal} {JPhys Energy}\ } (\bibinfo {year} {2024})}\BibitemShut {NoStop}%
\bibitem [{\citenamefont {Zhou}\ \emph {et~al.}(2025)\citenamefont {Zhou}, \citenamefont {Li}, \citenamefont {Huang},\ and\ \citenamefont {Chen}}]{zhou2025one}%
  \BibitemOpen
  \bibfield  {author} {\bibinfo {author} {\bibfnamefont {J.}~\bibnamefont {Zhou}}, \bibinfo {author} {\bibfnamefont {X.}~\bibnamefont {Li}}, \bibinfo {author} {\bibfnamefont {M.}~\bibnamefont {Huang}},\ and\ \bibinfo {author} {\bibfnamefont {S.}~\bibnamefont {Chen}},\ }\href@noop {} {\bibfield  {journal} {\bibinfo  {journal} {Phys. Rev. B}\ }\textbf {\bibinfo {volume} {112}},\ \bibinfo {pages} {235205} (\bibinfo {year} {2025})}\BibitemShut {NoStop}%
\bibitem [{\citenamefont {Yang}\ \emph {et~al.}(2025)\citenamefont {Yang}, \citenamefont {Liu}, \citenamefont {Zhang}, \citenamefont {Huang}, \citenamefont {Novoselov},\ and\ \citenamefont {Shen}}]{yang2025modeling}%
  \BibitemOpen
  \bibfield  {author} {\bibinfo {author} {\bibfnamefont {Z.}~\bibnamefont {Yang}}, \bibinfo {author} {\bibfnamefont {X.}~\bibnamefont {Liu}}, \bibinfo {author} {\bibfnamefont {X.}~\bibnamefont {Zhang}}, \bibinfo {author} {\bibfnamefont {P.}~\bibnamefont {Huang}}, \bibinfo {author} {\bibfnamefont {K.~S.}\ \bibnamefont {Novoselov}},\ and\ \bibinfo {author} {\bibfnamefont {L.}~\bibnamefont {Shen}},\ }\href@noop {} {\bibfield  {journal} {\bibinfo  {journal} {npj Comput. Mater.}\ }\textbf {\bibinfo {volume} {11}},\ \bibinfo {pages} {229} (\bibinfo {year} {2025})}\BibitemShut {NoStop}%
\bibitem [{\citenamefont {Turiansky}\ \emph {et~al.}(2025)\citenamefont {Turiansky}, \citenamefont {Lyons},\ and\ \citenamefont {Bernstein}}]{turiansky2025machinelearningphononspectra}%
  \BibitemOpen
  \bibfield  {author} {\bibinfo {author} {\bibfnamefont {M.~E.}\ \bibnamefont {Turiansky}}, \bibinfo {author} {\bibfnamefont {J.~L.}\ \bibnamefont {Lyons}},\ and\ \bibinfo {author} {\bibfnamefont {N.}~\bibnamefont {Bernstein}},\ }\href@noop {} {\bibfield  {journal} {\bibinfo  {journal} {arXiv}\ } (\bibinfo {year} {2025})},\ \bibinfo {note} {arxiv:2508.09113}\BibitemShut {NoStop}%
\bibitem [{\citenamefont {Mosquera-Lois}\ \emph {et~al.}(2025)\citenamefont {Mosquera-Lois}, \citenamefont {Klarbring},\ and\ \citenamefont {Walsh}}]{mosquera2025point}%
  \BibitemOpen
  \bibfield  {author} {\bibinfo {author} {\bibfnamefont {I.}~\bibnamefont {Mosquera-Lois}}, \bibinfo {author} {\bibfnamefont {J.}~\bibnamefont {Klarbring}},\ and\ \bibinfo {author} {\bibfnamefont {A.}~\bibnamefont {Walsh}},\ }\href@noop {} {\bibfield  {journal} {\bibinfo  {journal} {Chem. Sci.}\ }\textbf {\bibinfo {volume} {16}},\ \bibinfo {pages} {8878} (\bibinfo {year} {2025})}\BibitemShut {NoStop}%
\bibitem [{\citenamefont {Mosquera-Lois}\ and\ \citenamefont {Walsh}(2025)}]{mosquera2025dynamic}%
  \BibitemOpen
  \bibfield  {author} {\bibinfo {author} {\bibfnamefont {I.}~\bibnamefont {Mosquera-Lois}}\ and\ \bibinfo {author} {\bibfnamefont {A.}~\bibnamefont {Walsh}},\ }\href@noop {} {\bibfield  {journal} {\bibinfo  {journal} {PRX Energy}\ }\textbf {\bibinfo {volume} {4}},\ \bibinfo {pages} {043008} (\bibinfo {year} {2025})}\BibitemShut {NoStop}%
\bibitem [{\citenamefont {Rahman}\ and\ \citenamefont {Mannodi-Kanakkithodi}(2025)}]{rahman2025defect}%
  \BibitemOpen
  \bibfield  {author} {\bibinfo {author} {\bibfnamefont {M.~H.}\ \bibnamefont {Rahman}}\ and\ \bibinfo {author} {\bibfnamefont {A.}~\bibnamefont {Mannodi-Kanakkithodi}},\ }\href@noop {} {\bibfield  {journal} {\bibinfo  {journal} {arXiv}\ } (\bibinfo {year} {2025})},\ \bibinfo {note} {arXiv:2510.23514}\BibitemShut {NoStop}%
\bibitem [{\citenamefont {Cheng}(2025)}]{cheng2025latent}%
  \BibitemOpen
  \bibfield  {author} {\bibinfo {author} {\bibfnamefont {B.}~\bibnamefont {Cheng}},\ }\href@noop {} {\bibfield  {journal} {\bibinfo  {journal} {npj Comput. Mater.}\ }\textbf {\bibinfo {volume} {11}},\ \bibinfo {pages} {80} (\bibinfo {year} {2025})}\BibitemShut {NoStop}%
\bibitem [{\citenamefont {Ko}\ \emph {et~al.}(2021)\citenamefont {Ko}, \citenamefont {Finkler}, \citenamefont {Goedecker},\ and\ \citenamefont {Behler}}]{ko2021fourth}%
  \BibitemOpen
  \bibfield  {author} {\bibinfo {author} {\bibfnamefont {T.~W.}\ \bibnamefont {Ko}}, \bibinfo {author} {\bibfnamefont {J.~A.}\ \bibnamefont {Finkler}}, \bibinfo {author} {\bibfnamefont {S.}~\bibnamefont {Goedecker}},\ and\ \bibinfo {author} {\bibfnamefont {J.}~\bibnamefont {Behler}},\ }\href@noop {} {\bibfield  {journal} {\bibinfo  {journal} {Nat. Commun.}\ }\textbf {\bibinfo {volume} {12}},\ \bibinfo {pages} {398} (\bibinfo {year} {2021})}\BibitemShut {NoStop}%
\bibitem [{\citenamefont {Vondr{\'a}k}\ \emph {et~al.}(2026)\citenamefont {Vondr{\'a}k}, \citenamefont {Baldwin}, \citenamefont {Cs{\'a}nyi}, \citenamefont {Reuter},\ and\ \citenamefont {Margraf}}]{vondrak2026integrating}%
  \BibitemOpen
  \bibfield  {author} {\bibinfo {author} {\bibfnamefont {M.}~\bibnamefont {Vondr{\'a}k}}, \bibinfo {author} {\bibfnamefont {W.~J.}\ \bibnamefont {Baldwin}}, \bibinfo {author} {\bibfnamefont {G.}~\bibnamefont {Cs{\'a}nyi}}, \bibinfo {author} {\bibfnamefont {K.}~\bibnamefont {Reuter}},\ and\ \bibinfo {author} {\bibfnamefont {J.~T.}\ \bibnamefont {Margraf}},\ }\href@noop {} {\bibfield  {journal} {\bibinfo  {journal} {ChemRxiv}\ } (\bibinfo {year} {2026})},\ \bibinfo {note} {chemrxiv:15000377}\BibitemShut {NoStop}%
\bibitem [{\citenamefont {Batatia}\ \emph {et~al.}(2026)\citenamefont {Batatia}, \citenamefont {Baldwin}, \citenamefont {Kuryla}, \citenamefont {Hart}, \citenamefont {Kasoar}, \citenamefont {Elena}, \citenamefont {Moore}, \citenamefont {Gawkowski}, \citenamefont {Shi}, \citenamefont {Kapil} \emph {et~al.}}]{batatia2026mace}%
  \BibitemOpen
  \bibfield  {author} {\bibinfo {author} {\bibfnamefont {I.}~\bibnamefont {Batatia}}, \bibinfo {author} {\bibfnamefont {W.~J.}\ \bibnamefont {Baldwin}}, \bibinfo {author} {\bibfnamefont {D.}~\bibnamefont {Kuryla}}, \bibinfo {author} {\bibfnamefont {J.}~\bibnamefont {Hart}}, \bibinfo {author} {\bibfnamefont {E.}~\bibnamefont {Kasoar}}, \bibinfo {author} {\bibfnamefont {A.~M.}\ \bibnamefont {Elena}}, \bibinfo {author} {\bibfnamefont {H.}~\bibnamefont {Moore}}, \bibinfo {author} {\bibfnamefont {M.~J.}\ \bibnamefont {Gawkowski}}, \bibinfo {author} {\bibfnamefont {B.~X.}\ \bibnamefont {Shi}}, \bibinfo {author} {\bibfnamefont {V.}~\bibnamefont {Kapil}}, \emph {et~al.},\ }\href@noop {} {\bibfield  {journal} {\bibinfo  {journal} {arXiv}\ } (\bibinfo {year} {2026})},\ \bibinfo {note} {arXiv:2602.19411}\BibitemShut {NoStop}%
\bibitem [{\citenamefont {Baldwin}\ \emph {et~al.}(2026)\citenamefont {Baldwin}, \citenamefont {Batatia}, \citenamefont {Vondr{\'a}k}, \citenamefont {Margraf},\ and\ \citenamefont {Cs{\'a}nyi}}]{baldwin2026design}%
  \BibitemOpen
  \bibfield  {author} {\bibinfo {author} {\bibfnamefont {W.~J.}\ \bibnamefont {Baldwin}}, \bibinfo {author} {\bibfnamefont {I.}~\bibnamefont {Batatia}}, \bibinfo {author} {\bibfnamefont {M.}~\bibnamefont {Vondr{\'a}k}}, \bibinfo {author} {\bibfnamefont {J.~T.}\ \bibnamefont {Margraf}},\ and\ \bibinfo {author} {\bibfnamefont {G.}~\bibnamefont {Cs{\'a}nyi}},\ }\href@noop {} {\bibfield  {journal} {\bibinfo  {journal} {arXiv}\ } (\bibinfo {year} {2026})},\ \bibinfo {note} {arXiv:2603.14700}\BibitemShut {NoStop}%
\bibitem [{\citenamefont {Wang}\ \emph {et~al.}(2014)\citenamefont {Wang}, \citenamefont {Li},\ and\ \citenamefont {Truhlar}}]{wang2014modeling}%
  \BibitemOpen
  \bibfield  {author} {\bibinfo {author} {\bibfnamefont {B.}~\bibnamefont {Wang}}, \bibinfo {author} {\bibfnamefont {S.~L.}\ \bibnamefont {Li}},\ and\ \bibinfo {author} {\bibfnamefont {D.~G.}\ \bibnamefont {Truhlar}},\ }\href@noop {} {\bibfield  {journal} {\bibinfo  {journal} {J. Chem. Theory Comput.}\ }\textbf {\bibinfo {volume} {10}},\ \bibinfo {pages} {5640} (\bibinfo {year} {2014})}\BibitemShut {NoStop}%
\bibitem [{\citenamefont {Toukmaji}\ and\ \citenamefont {Board~Jr}(1996)}]{toukmaji1996ewald}%
  \BibitemOpen
  \bibfield  {author} {\bibinfo {author} {\bibfnamefont {A.~Y.}\ \bibnamefont {Toukmaji}}\ and\ \bibinfo {author} {\bibfnamefont {J.~A.}\ \bibnamefont {Board~Jr}},\ }\href@noop {} {\bibfield  {journal} {\bibinfo  {journal} {Comput. Phys. Commun.}\ }\textbf {\bibinfo {volume} {95}},\ \bibinfo {pages} {73} (\bibinfo {year} {1996})}\BibitemShut {NoStop}%
\bibitem [{\citenamefont {Wood}\ \emph {et~al.}(2025)\citenamefont {Wood}, \citenamefont {Dzamba}, \citenamefont {Fu}, \citenamefont {Gao}, \citenamefont {Shuaibi}, \citenamefont {Barroso-Luque}, \citenamefont {Abdelmaqsoud}, \citenamefont {Gharakhanyan}, \citenamefont {Kitchin}, \citenamefont {Levine} \emph {et~al.}}]{wood2025family}%
  \BibitemOpen
  \bibfield  {author} {\bibinfo {author} {\bibfnamefont {B.~M.}\ \bibnamefont {Wood}}, \bibinfo {author} {\bibfnamefont {M.}~\bibnamefont {Dzamba}}, \bibinfo {author} {\bibfnamefont {X.}~\bibnamefont {Fu}}, \bibinfo {author} {\bibfnamefont {M.}~\bibnamefont {Gao}}, \bibinfo {author} {\bibfnamefont {M.}~\bibnamefont {Shuaibi}}, \bibinfo {author} {\bibfnamefont {L.}~\bibnamefont {Barroso-Luque}}, \bibinfo {author} {\bibfnamefont {K.}~\bibnamefont {Abdelmaqsoud}}, \bibinfo {author} {\bibfnamefont {V.}~\bibnamefont {Gharakhanyan}}, \bibinfo {author} {\bibfnamefont {J.~R.}\ \bibnamefont {Kitchin}}, \bibinfo {author} {\bibfnamefont {D.~S.}\ \bibnamefont {Levine}}, \emph {et~al.},\ }\href@noop {} {\bibfield  {journal} {\bibinfo  {journal} {arXiv}\ } (\bibinfo {year} {2025})},\ \bibinfo {note} {arXiv:2506.23971}\BibitemShut {NoStop}%
\bibitem [{\citenamefont {Shimizu}\ \emph {et~al.}(2022)\citenamefont {Shimizu}, \citenamefont {Dou}, \citenamefont {Arguelles}, \citenamefont {Moriya}, \citenamefont {Minamitani},\ and\ \citenamefont {Watanabe}}]{shimizu2022using}%
  \BibitemOpen
  \bibfield  {author} {\bibinfo {author} {\bibfnamefont {K.}~\bibnamefont {Shimizu}}, \bibinfo {author} {\bibfnamefont {Y.}~\bibnamefont {Dou}}, \bibinfo {author} {\bibfnamefont {E.~F.}\ \bibnamefont {Arguelles}}, \bibinfo {author} {\bibfnamefont {T.}~\bibnamefont {Moriya}}, \bibinfo {author} {\bibfnamefont {E.}~\bibnamefont {Minamitani}},\ and\ \bibinfo {author} {\bibfnamefont {S.}~\bibnamefont {Watanabe}},\ }\href@noop {} {\bibfield  {journal} {\bibinfo  {journal} {Phys. Rev.B}\ }\textbf {\bibinfo {volume} {106}},\ \bibinfo {pages} {054108} (\bibinfo {year} {2022})}\BibitemShut {NoStop}%
\bibitem [{\citenamefont {Kiyohara}\ \emph {et~al.}(2025)\citenamefont {Kiyohara}, \citenamefont {Shibui}, \citenamefont {Bae},\ and\ \citenamefont {Kumagai}}]{kiyohara2025machine}%
  \BibitemOpen
  \bibfield  {author} {\bibinfo {author} {\bibfnamefont {S.}~\bibnamefont {Kiyohara}}, \bibinfo {author} {\bibfnamefont {C.}~\bibnamefont {Shibui}}, \bibinfo {author} {\bibfnamefont {S.}~\bibnamefont {Bae}},\ and\ \bibinfo {author} {\bibfnamefont {Y.}~\bibnamefont {Kumagai}},\ }\href@noop {} {\bibfield  {journal} {\bibinfo  {journal} {Phys. Rev. Lett.}\ }\textbf {\bibinfo {volume} {135}},\ \bibinfo {pages} {246101} (\bibinfo {year} {2025})}\BibitemShut {NoStop}%
\bibitem [{\citenamefont {Wang}\ \emph {et~al.}(2024{\natexlab{b}})\citenamefont {Wang}, \citenamefont {Kavanagh}, \citenamefont {Scanlon},\ and\ \citenamefont {Walsh}}]{wang2024upper}%
  \BibitemOpen
  \bibfield  {author} {\bibinfo {author} {\bibfnamefont {X.}~\bibnamefont {Wang}}, \bibinfo {author} {\bibfnamefont {S.~R.}\ \bibnamefont {Kavanagh}}, \bibinfo {author} {\bibfnamefont {D.~O.}\ \bibnamefont {Scanlon}},\ and\ \bibinfo {author} {\bibfnamefont {A.}~\bibnamefont {Walsh}},\ }\href@noop {} {\bibfield  {journal} {\bibinfo  {journal} {Joule}\ }\textbf {\bibinfo {volume} {8}},\ \bibinfo {pages} {2105} (\bibinfo {year} {2024}{\natexlab{b}})}\BibitemShut {NoStop}%
\bibitem [{\citenamefont {Fan}\ \emph {et~al.}(2026)\citenamefont {Fan}, \citenamefont {Tang}, \citenamefont {Berger}, \citenamefont {Berger}, \citenamefont {Fransson}, \citenamefont {Xu}, \citenamefont {Yan}, \citenamefont {Liu}, \citenamefont {Song}, \citenamefont {Dong} \emph {et~al.}}]{fan2026qnep}%
  \BibitemOpen
  \bibfield  {author} {\bibinfo {author} {\bibfnamefont {Z.}~\bibnamefont {Fan}}, \bibinfo {author} {\bibfnamefont {B.}~\bibnamefont {Tang}}, \bibinfo {author} {\bibfnamefont {E.}~\bibnamefont {Berger}}, \bibinfo {author} {\bibfnamefont {E.}~\bibnamefont {Berger}}, \bibinfo {author} {\bibfnamefont {E.}~\bibnamefont {Fransson}}, \bibinfo {author} {\bibfnamefont {K.}~\bibnamefont {Xu}}, \bibinfo {author} {\bibfnamefont {Z.}~\bibnamefont {Yan}}, \bibinfo {author} {\bibfnamefont {Z.}~\bibnamefont {Liu}}, \bibinfo {author} {\bibfnamefont {Z.}~\bibnamefont {Song}}, \bibinfo {author} {\bibfnamefont {H.}~\bibnamefont {Dong}}, \emph {et~al.},\ }\href@noop {} {\bibfield  {journal} {\bibinfo  {journal} {arXiv}\ } (\bibinfo {year} {2026})},\ \bibinfo {note} {arXiv:2601.19034}\BibitemShut {NoStop}%
\bibitem [{\citenamefont {Mori-S{\'a}nchez}\ \emph {et~al.}(2008)\citenamefont {Mori-S{\'a}nchez}, \citenamefont {Cohen},\ and\ \citenamefont {Yang}}]{mori2008localization}%
  \BibitemOpen
  \bibfield  {author} {\bibinfo {author} {\bibfnamefont {P.}~\bibnamefont {Mori-S{\'a}nchez}}, \bibinfo {author} {\bibfnamefont {A.~J.}\ \bibnamefont {Cohen}},\ and\ \bibinfo {author} {\bibfnamefont {W.}~\bibnamefont {Yang}},\ }\href@noop {} {\bibfield  {journal} {\bibinfo  {journal} {Phys. Rev. Lett.}\ }\textbf {\bibinfo {volume} {100}},\ \bibinfo {pages} {146401} (\bibinfo {year} {2008})}\BibitemShut {NoStop}%
\bibitem [{\citenamefont {Squires}\ \emph {et~al.}(2026)\citenamefont {Squires}, \citenamefont {Kavanagh}, \citenamefont {Walsh},\ and\ \citenamefont {Scanlon}}]{squires2025guidelines}%
  \BibitemOpen
  \bibfield  {author} {\bibinfo {author} {\bibfnamefont {A.~G.}\ \bibnamefont {Squires}}, \bibinfo {author} {\bibfnamefont {S.}~\bibnamefont {Kavanagh}}, \bibinfo {author} {\bibfnamefont {A.}~\bibnamefont {Walsh}},\ and\ \bibinfo {author} {\bibfnamefont {D.~O.}\ \bibnamefont {Scanlon}},\ }\href@noop {} {\bibfield  {journal} {\bibinfo  {journal} {Nat. Rev. Mater.}\ } (\bibinfo {year} {2026})}\BibitemShut {NoStop}%
\bibitem [{\citenamefont {Ko}\ \emph {et~al.}(2025)\citenamefont {Ko}, \citenamefont {Liu}, \citenamefont {Mishra}, \citenamefont {Yu}, \citenamefont {Qi},\ and\ \citenamefont {Ong}}]{ko2025fast}%
  \BibitemOpen
  \bibfield  {author} {\bibinfo {author} {\bibfnamefont {T.~W.}\ \bibnamefont {Ko}}, \bibinfo {author} {\bibfnamefont {R.}~\bibnamefont {Liu}}, \bibinfo {author} {\bibfnamefont {A.~R.}\ \bibnamefont {Mishra}}, \bibinfo {author} {\bibfnamefont {Z.}~\bibnamefont {Yu}}, \bibinfo {author} {\bibfnamefont {J.}~\bibnamefont {Qi}},\ and\ \bibinfo {author} {\bibfnamefont {S.~P.}\ \bibnamefont {Ong}},\ }\href@noop {} {\bibfield  {journal} {\bibinfo  {journal} {arXiv preprint arXiv:2511.07249}\ } (\bibinfo {year} {2025})}\BibitemShut {NoStop}%
\bibitem [{\citenamefont {Mosquera-Lois}\ \emph {et~al.}(2023)\citenamefont {Mosquera-Lois}, \citenamefont {Kavanagh}, \citenamefont {Walsh},\ and\ \citenamefont {Scanlon}}]{mosquera2023identifying}%
  \BibitemOpen
  \bibfield  {author} {\bibinfo {author} {\bibfnamefont {I.}~\bibnamefont {Mosquera-Lois}}, \bibinfo {author} {\bibfnamefont {S.~R.}\ \bibnamefont {Kavanagh}}, \bibinfo {author} {\bibfnamefont {A.}~\bibnamefont {Walsh}},\ and\ \bibinfo {author} {\bibfnamefont {D.~O.}\ \bibnamefont {Scanlon}},\ }\href@noop {} {\bibfield  {journal} {\bibinfo  {journal} {npj Comput. Mater.}\ }\textbf {\bibinfo {volume} {9}},\ \bibinfo {pages} {25} (\bibinfo {year} {2023})}\BibitemShut {NoStop}%
\bibitem [{\citenamefont {Alkauskas}\ \emph {et~al.}(2014)\citenamefont {Alkauskas}, \citenamefont {Yan},\ and\ \citenamefont {Van~de Walle}}]{alkauskas2014first}%
  \BibitemOpen
  \bibfield  {author} {\bibinfo {author} {\bibfnamefont {A.}~\bibnamefont {Alkauskas}}, \bibinfo {author} {\bibfnamefont {Q.}~\bibnamefont {Yan}},\ and\ \bibinfo {author} {\bibfnamefont {C.~G.}\ \bibnamefont {Van~de Walle}},\ }\href@noop {} {\bibfield  {journal} {\bibinfo  {journal} {Phys. Rev. B}\ }\textbf {\bibinfo {volume} {90}},\ \bibinfo {pages} {075202} (\bibinfo {year} {2014})}\BibitemShut {NoStop}%
\bibitem [{\citenamefont {Hainer}\ \emph {et~al.}(2025)\citenamefont {Hainer}, \citenamefont {Berger}, \citenamefont {Berger}, \citenamefont {Hildeberg}, \citenamefont {Erhart},\ and\ \citenamefont {Wiktor}}]{hainer2025thermalstabilizationdefectcharge}%
  \BibitemOpen
  \bibfield  {author} {\bibinfo {author} {\bibfnamefont {T.}~\bibnamefont {Hainer}}, \bibinfo {author} {\bibfnamefont {E.}~\bibnamefont {Berger}}, \bibinfo {author} {\bibfnamefont {E.}~\bibnamefont {Berger}}, \bibinfo {author} {\bibfnamefont {O.}~\bibnamefont {Hildeberg}}, \bibinfo {author} {\bibfnamefont {P.}~\bibnamefont {Erhart}},\ and\ \bibinfo {author} {\bibfnamefont {J.}~\bibnamefont {Wiktor}},\ }\href@noop {} {\bibfield  {journal} {\bibinfo  {journal} {arXiv}\ } (\bibinfo {year} {2025})},\ \bibinfo {note} {arxiv:2512.15463}\BibitemShut {NoStop}%
\bibitem [{\citenamefont {Linder\"{a}lv}\ \emph {et~al.}(2025)\citenamefont {Linder\"{a}lv}, \citenamefont {\"{O}sterbacka}, \citenamefont {Wiktor},\ and\ \citenamefont {Erhart}}]{Linderlv2025}%
  \BibitemOpen
  \bibfield  {author} {\bibinfo {author} {\bibfnamefont {C.}~\bibnamefont {Linder\"{a}lv}}, \bibinfo {author} {\bibfnamefont {N.}~\bibnamefont {\"{O}sterbacka}}, \bibinfo {author} {\bibfnamefont {J.}~\bibnamefont {Wiktor}},\ and\ \bibinfo {author} {\bibfnamefont {P.}~\bibnamefont {Erhart}},\ }\bibfield  {journal} {\bibinfo  {journal} {npj Comput. Mater.}\ }\textbf {\bibinfo {volume} {11}},\ \href {https://doi.org/10.1038/s41524-025-01565-x} {10.1038/s41524-025-01565-x} (\bibinfo {year} {2025})\BibitemShut {NoStop}%
\bibitem [{\citenamefont {Tyagi}\ \emph {et~al.}(2025)\citenamefont {Tyagi}, \citenamefont {Pols}, \citenamefont {Brocks},\ and\ \citenamefont {Tao}}]{Tyagi2025}%
  \BibitemOpen
  \bibfield  {author} {\bibinfo {author} {\bibfnamefont {V.}~\bibnamefont {Tyagi}}, \bibinfo {author} {\bibfnamefont {M.}~\bibnamefont {Pols}}, \bibinfo {author} {\bibfnamefont {G.}~\bibnamefont {Brocks}},\ and\ \bibinfo {author} {\bibfnamefont {S.}~\bibnamefont {Tao}},\ }\href {https://doi.org/10.1021/acs.jpclett.5c01139} {\bibfield  {journal} {\bibinfo  {journal} {J. Phys. Chem. Lett.}\ }\textbf {\bibinfo {volume} {16}},\ \bibinfo {pages} {5153–5159} (\bibinfo {year} {2025})}\BibitemShut {NoStop}%
\end{thebibliography}%

\clearpage
\onecolumngrid

\includepdf[
    pages={{},-},
    pagecommand={}
]{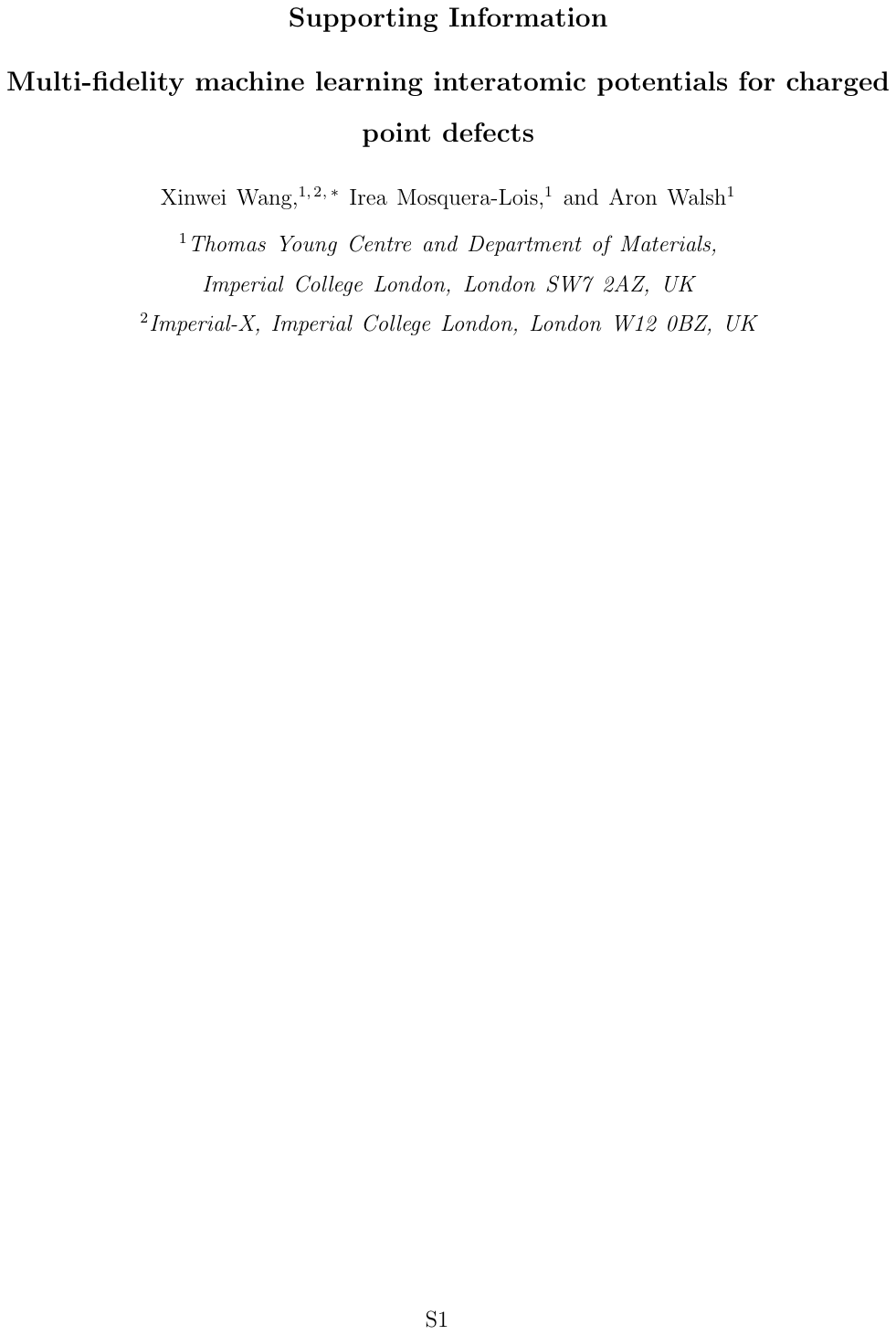}

\end{document}